\documentclass[useAMS,usenatbib,usegraphicx]{mn2e}
\usepackage{epsf}

\title[The largest Einstein radius]
{What is the largest Einstein radius in the universe?}

\author[M. Oguri and R. D. Blandford]
{Masamune Oguri$^{1}$\thanks{E-mail: oguri@slac.stanford.edu} 
and Roger D. Blandford$^{1}$\\
$^1$Kavli Institute for Particle Astrophysics and Cosmology, 
Stanford University, 2575 Sand Hill Road, Menlo Park, CA
94025, USA.} 

\begin{document}

\date{\today}

\voffset- .5in

\pagerange{\pageref{firstpage}--\pageref{lastpage}} \pubyear{}

\maketitle

\label{firstpage}

\begin{abstract}
The Einstein radius plays a central role in lens studies as it
characterises the strength of gravitational lensing. In particular,
the distribution of Einstein radii near the upper cutoff should probe the
probability distribution of the largest mass concentrations
in the universe. Adopting a triaxial halo model, we compute 
expected distributions of large Einstein radii. To assess the cosmic
variance, we generate a number of Monte-Carlo realisations of all-sky
catalogues of massive clusters.  We find that the expected largest Einstein
radius in the universe is sensitive to parameters characterising the
cosmological model, especially $\sigma_8$: for a source redshift of
unity, they are $42{}^{+9}_{-7}$, $35{}^{+8}_{-6}$, and
$54{}^{+12}_{-7}$ arcseconds (errors denote $1\sigma$ cosmic
variance), assuming best-fit cosmological parameters of the Wilkinson
Microwave Anisotropy Probe five-year (WMAP5), three-year (WMAP3) and
one-year (WMAP1) data, respectively. These values are broadly
consistent with current observations given their incompleteness.  The
mass of the largest lens cluster can be as small as
$\sim10^{15}M_\odot$. For the same source redshift, we expect in
all-sky $\sim 35$ (WMAP5), $\sim 15$ (WMAP3), and $\sim 150$ (WMAP1)
clusters that have Einstein radii larger than $20''$. For a larger
source redshift of 7, the largest Einstein radii grow approximately
twice as large. Whilst the values of the largest Einstein radii are
almost unaffected by the level of the primordial  non-Gaussianity
currently of interest, the  measurement of the abundance of moderately
large lens clusters should probe non-Gaussianity competitively with
cosmic microwave background experiments, but only if other
cosmological parameters are well-measured. These semi-analytic
predictions are based on a rather simple representation of clusters,
and hence calibrating them with $N$-body simulations will help to
improve the accuracy. We also find that these ``superlens'' clusters
constitute a highly biased population. For instance, a substantial
fraction of these superlens clusters have major axes preferentially
aligned with the line-of-sight. As a consequence, the projected mass
distributions of the clusters are rounder by an ellipticity of $\sim
0.2$ and have $\sim 40\%-60\%$ larger concentrations compared with
typical clusters with similar redshifts and masses. We argue that the
large concentration measured in A1689 is consistent with our model
prediction at the $1.2\sigma$ level. A combined analysis of several
clusters will be needed to see whether or not the observed
concentrations conflict with predictions of the flat
$\Lambda$-dominated cold dark matter model. 
\end{abstract}


\section{Introduction}
\label{sec:intro}

In the standard Cold Dark Matter (CDM) model, which for
these purposes we shall assume includes the presence of a cosmological
constant and a flat spatial geometry, structure grows hierarchically
from small objects that merge together to form larger objects
(hereafter F$\Lambda$CDM). Strong
gravitational lensing by massive clusters of galaxies is one of the
most important tests of this model in the sense that it probes the
rarest high density peaks in the universe. For instance, the CDM
model predicts wide-angle lensing events, on scales as large as
several tens arcseconds, due to massive clusters
\citep[e.g.,][]{turner84,narayan84,narayan88,wambsganss95}. This has
been broadly verified by the discovery of many lensed background galaxies
\citep[e.g.,][]{lefevre94,luppino99,gladders03,zaritsky03,sand05,hennawi08}
or quasars \citep{inada03,inada06}. Quantitative comparisons of
expected lensing rates in the F$\Lambda$CDM model and observed numbers
of lenses should serve as an important test of our understanding of
the universe.  

A possible simple test of the F$\Lambda$CDM model is the statistics of
Einstein radii, particularly near the upper cutoff. The Einstein
radius is a characteristic scale of strong lensing and is related
mainly to the aperture mass it encloses. Therefore it is expected that
the largest Einstein radii in the universe probe the structure and
abundance of the most massive clusters. This enables a test of the
F$\Lambda$CDM model at the very tail of the halo distribution. An
advantage of this test is the simple and straightforward determination
of the Einstein radius in observations and its correspondence to
identify large lenses. 

Lensing properties of massive clusters have mainly been studied using 
ray-tracing in $N$-body simulations \citep[e.g.,][]{wambsganss95,
wambsganss04,bartelmann95,bartelmann98,meneghetti03a,meneghetti05,
dalal04,ho05,li05,horesh05,hennawi07a,hennawi07b,hilbert07,hilbert08}. 
Whilst the numerical approach allows one to take account of the full
complexity of lens potentials, it is often computationally expensive
to perform large box-size simulations which retain enough particles in
each halo for strong lensing studies. In particular reliable
predictions for the rarest lensing events in the universe require
many realisations of such high-resolution Hubble-size simulations in
order to estimate the effect of the cosmic variance. This is
impractical with current computational capabilities.  

The complementary semi-analytic approaches often invoked simple spherically
symmetric mass profiles, for calculational reasons 
\citep[e.g.,][]{maoz97,hamana97,molikawa99,cooray99,wyithe01,takahashi01,
kochanek01,keeton01,li03,lopes04,huterer04,kuhlen04,chen04,chen05,oguri06}. 
A more advanced calculation adopted an ellipsoid for projected
cluster mass distributions 
\citep{meneghetti03a,fedeli07a,fedeli07b,fedeli08}.  
However, because of the triaxial nature of F$\Lambda$CDM haloes
\citep[e.g.,][]{jing02,allgood06,shaw06,hayashi07} the lensing
properties of individual clusters vary drastically as a function of
viewing angle \citep[e.g.,][]{dalal04,hennawi07b}, resulting in the
significant increase of average lensing efficiencies due to halo
triaxiality. This indicates that any analytic models of cluster
lensing should take proper account of triaxiality for reliable
theoretical predictions, as is done in several papers
\citep{oguri03,oguri04,minor08}.  

In this paper, we take a semi-analytic approach to predict the
largest Einstein radius in all-sky survey, based on the F$\Lambda$CDM
model. We invoke an analytic mass function of dark haloes to generate
a catalogue of massive clusters with the Monte-Carlo method. The shape
of each halo is assumed to be triaxial, and the projection along
random directions is considered. This Monte-Carlo approach allows us
to  evaluate the range of the largest Einstein radii due to 
cosmic variance. We also characterise such ``superlens'' clusters,
i.e., clusters which produce widest-angle lensing, to see how
``unusual'' are these clusters.   

These issues are well illustrated by detailed observations of the
largest Einstein radius known to data, which may conflict with the
F$\Lambda$CDM model. The lensing data of A1689, 
one of the best-studied clusters to date, suggest that the
mass profile is apparently more centrally concentrated
\citep{broadhurst05a,broadhurst08a} than the F$\Lambda$CDM
prediction \citep[e.g.,][]{neto07,duffy08,maccio08},  although the
exact degree of concentration is somewhat controversial
\citep[e.g.,][]{halkola06,medezinski07,limousin07,comerford07,umetsu08}. 
It has been argued that a part of discrepancy can be explained by 
halo triaxiality \citep{oguri05a,gavazzi05,hennawi07b,corless07,corless08} 
or the projection of the secondary mass peak along the line-of-sight
\citep{andersson04,lokas06,king07}, suggesting the importance of careful
statistical studies with the selection effect taken into account. 
Indeed, it should be pointed out that a weak lensing analysis of stacked
clusters of lesser mass does not exhibit the high concentration
problem \citep{johnston08,mandelbaum08}.

We believe that our predictions will be helpful for interpreting
surveys of distant ($z\ga 6$) galaxies near critical curves of massive
clusters
\citep{ellis01,hu02,kneib04,richard06,richard08,stark07,willis08,bowens08}.  
The survey area of this technique is simply proportional to the square of
the Einstein radius, thus clusters with very large Einstein radii are
thought to be the best sites to conduct this search. Our calculations
should provide a useful guidance to discover such giant lens clusters. 

This paper is organised as follows. We describe our theoretical model
in Section~\ref{sec:mc}. Predictions for the largest Einstein radius and
the abundance of large lens clusters are shown in Section~\ref{sec:largest}
and Section~\ref{sec:num}, respectively.  Section~\ref{sec:ng}
includes the effect of primordial non-Gaussianities. We discuss the
results in Section~\ref{sec:dis}, and give our conclusion in
Section~\ref{sec:conc}.  
Throughout the paper, we consider three cosmological parameter sets
obtained from the {\it Wilkinson Microwave Anisotropy Probe
  (WMAP)}, mainly to show how sensitive our results are to
cosmological parameters. These are the best-fit parameter sets from
the WMAP one-year data \citep[WMAP1;][]{spergel03}, ($\Omega_M$,
$\Omega_b$, $\Omega_\Lambda$, $h$, $n_s$, $\sigma_8$)$=$(0.270, 0.046,
0.730, 0.72, 0.99, 0.9), WMAP three-year data
\citep[WMAP3;][]{spergel07}, (0.238, 0.042, 0.762,
0.732, 0.958, 0.761), and WMAP five-year data
\citep[WMAP5;][]{dunkley08}, (0.258, 0.044, 0.742, 
0.719, 0.963, 0.796). The most important difference between these
models is the matter density and the normalisation of matter
fluctuations. Indeed, it has been shown that the smaller values of
$\Omega_M$ and $\sigma_8$ in WMAP3 resulted in much smaller number of
cluster-scale lenses compared with WMAP1 \citep[e.g.,][]{li06,li07}. 
Unless otherwise specified, we adopt the WMAP5 cosmology as our
fiducial cosmological model.    

\section{Monte-Carlo approach to the distribution of Einstein radii}
\label{sec:mc}
We compute the cosmological distribution of Einstein radii
semi-analytically using a Monte-Carlo technique. First, we randomly
generate a catalogue of massive dark haloes according to a mass
function. We assume a fitting formula derived by \citet{warren06} for
the mass function of dark haloes: 
\begin{equation}
\frac{dn}{dM}=
0.7234\left(\sigma_M^{-1.625}+0.2538\right)e^{-1.1982/\sigma_M^2}
\frac{\rho(z)}{M^2}\frac{d\ln\sigma^{-1}_M}{d\ln M},
\label{eq:mf}
\end{equation}
where $M$ is a halo mass and $\rho(z)$ is a mean comoving matter
density at redshift $z$. We calculate the linear density fluctuation
$\sigma_M$ from the approximated transfer function presented
by \citet{eisenstein98}. Throughout the paper we adopt the virial
mass $M=M_{\rm vir}$ which is defined such that the average density
inside a spherical region with mass $M_{\rm vir}$ becomes $\Delta(z)$
times the mean matter density of the universe; here $\Delta(z)$ is
computed using the spherical collapse model in the F$\Lambda$CDM
universe. For our fiducial cosmological model, $\Delta(0)\approx 370$,
$\Delta(0.3)\approx 270$, and $\Delta(1)\approx 200$ (see, e.g.,
\citealt{nakamura97}). In this paper we are interested 
in massive clusters with the 
masses $M\sim 10^{15}M_\odot$. For comparison, the mass of the Coma
cluster is $\sim 1.3\times 10^{15}M_\odot$ \citep{hughes89}, and that
of A1689 is $\sim 2.1\times10^{15}M_\odot$ \citep{umetsu08}.
With this mass function, the number of dark
haloes for each redshift and mass bin can be written as 

\begin{equation}
N=\frac{d^2N}{dzdM}\Delta z\Delta M=\Omega D_A(z)^2\frac{dr}{dz}
(1+z)^3\frac{dn}{dM}\Delta z\Delta M,
\end{equation}
with $D_A(z)$ and $dr/dz$ being the angular diameter distance and the
proper differential distance, respectively. Throughout the paper we
adopt the solid angle of $\Omega = 40,000$~${\rm deg^2}$ which roughly
corresponds to all-sky excluding the Galactic plane. A realisation of
dark haloes is then constructed by computing the expected mean number
of dark haloes for each bin adopting $\Delta z=0.01$ and $\Delta (\log
M)=0.02$, and generating an integer number from the Poisson
distribution with the mean. The Monte-Carlo catalogues are generated
in the range of cluster masses larger than the minimum mass $M_{\rm
  min}$. We adopt $M_{\rm min}=4\times 10^{14}M_\odot$ for WMAP1 and
$M_{\rm min}=2\times 10^{14}M_\odot$ for WMAP3 and WMAP5, which are
sufficiently small not to affect our results. On the other hand,
the maximum cluster masses for these cosmologies are $M_{\rm max}\sim
3-5\times 10^{15}M_\odot$ (see \S\ref{sec:clumz}). 

Each dark halo is assumed to have a triaxial shape. Following
\citet[][hereafter JS02]{jing02}, we model the density profile as
\begin{equation}
 \rho(R)=\frac{\delta_{\rm cd}\rho_{\rm crit}(z)}
{(R/R_0)(1+R/R_0)^{2}}\left\{\frac{1}{1+(R/R_t)^2}\right\}^2,
\label{eq:tnfw}
\end{equation}
\begin{equation}
 R^2\equiv c^2\left(\frac{x^2}{a^2}+\frac{y^2}{b^2}
+\frac{z^2}{c^2}\right)\;\;\;\;\;(a\leq b\leq c).
\label{eq:rdef}
\end{equation}
The model is a triaxial generalisation of the
\citet[][hereafter NFW]{navarro97} density profile. The concentration
parameter for this triaxial model is defined by $c_e\equiv R_e/R_0$,
where $R_e$ is determined such that the mean density within the
ellipsoid of the major axis radius $R_e$ is
$5\Delta(z)\left(c^2/ab\right)^{0.75}\rho(z)(1+z)^3$.  The 
characteristic density $\delta_{\rm cd}$ is then written in terms of
the concentration parameter. As suggested in JS02 we relate $R_e$
to the virial mass $M_{\rm vir}$ of the halo by adopting a 
relation $R_e/r_{\rm vir}=0.45$, where $r_{\rm vir}$  is spherical
virial radius computed from the virial mass. A change from JS02 is the
inclusion of a truncation term, $[1+(R/R_t)^2]^{-2}$, such that the
radial profile does not extend far beyond the virial radius
\citep[][see also \citealt{takada03}]{baltz08}.  We choose
$R_t=4r_{\rm vir}$ which can be translated into the truncation at
roughly twice the virial radius for massive haloes.  We note that the
truncation is introduced for haloes with very small $c_e$; in this 
situation masses outside the virial radii dominate the lens potentials
\citep[see][]{oguri04}, and thus the truncation is necessary to avoid 
such unrealistic situations. The truncation has a negligible effect on
the Einstein radii of most haloes.  

\begin{figure}
\begin{center}
 \includegraphics[width=0.95\hsize]{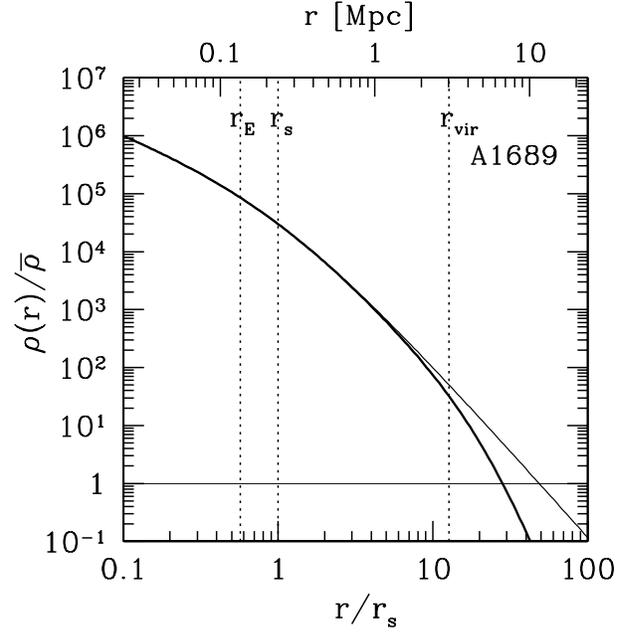}
\end{center}
\caption{The illustration of various characteristic scales for large
  cluster lenses. The best-fit spherical NFW density profile of A1689
  \citep{umetsu08} $\rho(r)$ normalised by the mean matter density
  of the universe at that redshift $\bar{\rho}$ is plotted as a
  function of the radius. Thick and thin lines show the NFW profiles
  with and without the truncation term (see eq. [\ref{eq:tnfw}]). The
  Einstein radius for the source redshift of unity, the scale radius of the
  spherical NFW profile, and the virial radius of the cluster are
  indicated by vertical dotted lines with labels of $r_{\rm E}$,
  $r_s$~($\sim R_0$), and  $r_{\rm vir}$, respectively. The total 
  mass of a sphere defined by the Einstein radius is $1.1\times
  10^{14}M_\odot$, and the cylindrical mass projected within the
  Einstein radius is $2.0\times 10^{14}M_\odot$, which should be
  compared with the virial mass (total mass inside $r_{\rm vir}$),
  $M_{\rm vir}=2.1\times10^{15}M_\odot$.  
\label{fig:nfwpro}}
\end{figure}

To give a rough idea of various length scales for massive lensing
clusters, in Figure \ref{fig:nfwpro} we plot the best-fit radial NFW
density profile of A1689 derived from lensing \citep{umetsu08}. The
Einstein radii of massive lensing clusters are typically $\sim 5$\% of
the virial radii $r_{\rm vir}$ which are a few Mpc for these clusters. 
The density at the Einstein radius is $\sim 10^5$ times more than the
mean matter density of the universe $\bar{\rho}$. The Figure also
indicates that our truncation of the NFW profile (see
eq. [\ref{eq:tnfw}]) only affects the radial density profile at $r\ga
r_{\rm vir}$.  The radial profile crosses the mean matter density
$\bar{\rho}$ at several times the virial radius. 

The axis ratio and concentration parameter for each halo are randomly
assigned according to the probability distribution functions (PDFs)
derived by JS02. Specifically the probability distributions for
triaxial axis ratios are given by 
\begin{equation}
 p(a/c)=\frac{1}{\sqrt{2\pi}\sigma_s}\exp\left[-\frac{\left(r_{\rm
 ac}-0.54\right)^2}{2\sigma_s^2}\right]\frac{dr_{\rm ac}}{d(a/c)},
\label{eq:p_a}
\end{equation}
\begin{equation}
 p(a/b|a/c)=\frac{3}{2(1-r_{\rm min})}\left[1-\left(\frac{2a/b-1-
 r_{\rm min}}{1-r_{\rm min}}\right)^2\right],
\label{eq:p_ab}
\end{equation}
\begin{equation}
r_{\rm ac}=\frac{a}{c}\left(\frac{M}{M_*}
\right)^{0.07[\Omega(z)]^{0.7}},
\end{equation}
\begin{equation}
r_{\rm min}={\rm Max}[a/c,0.5].
\end{equation}
The characteristic nonlinear mass $M_*$ is defined such that the
overdensity at that mass scale becomes $\delta_c=1.68$. The best-fit
value for the width of the axis ratio distribution $\sigma_s$ is
$\sigma_s=0.113$. Note that $p(a/b|a/c)=0$ 
for $a/b<r_{\rm min}$. On the other hand, the probability distribution
for the concentration parameter is well approximated by the log-normal
distribution: 
\begin{equation}
 p(c_e)=\frac{1}{\sqrt{2\pi}\sigma_c} \exp\left[
-\frac{(\ln c_e-\ln \bar{c}_e)^2}{2\sigma_c^2}\right]\frac{1}{c_e}, 
\label{eq:p_ce}
\end{equation}
with the width of the distribution $\sigma_c=0.3$. We include a
correlation between the axis ratio  and concentration parameter by
adopting the following form for the median concentration parameter
$\bar{c}_e$ \citep[see][]{oguri03}: 
\begin{equation}
 \bar{c}_e={\rm Max}[f_c,0.3]A_e\sqrt{\frac{\Delta(z_c)}
{\Delta(z)}}\left(\frac{1+z_c}{1+z}\right)^{3/2},
\label{eq:ce}
\end{equation}
\begin{equation}
 f_c=1.35\exp\left[-\left(\frac{0.3}{r_{\rm ac}}\right)^2\right],
\label{eq:fc}
\end{equation}
where $A_e=1.1$ for F$\Lambda$CDM. The collapse redshift $z_c$ of
a halo with mass $M$ is estimated by solving the following equation
involving the complementary error function:
\begin{equation}
{\rm erfc}\frac{\delta_c(z)-\delta_c(0)}{\sqrt{2(\sigma^2_{fM}
-\sigma^2_M)}}=\frac{1}{2}.
\label{eq:colz}
\end{equation}
Here the linear overdensity at redshift $z$ is
$\delta_c(z)=\delta_c/D(z)$ with $D(z)$ being the linear growth
rate, and $\sigma_{fM}$ and $\sigma_M$ are linear density fluctuations
for the mass scales of $fM$ and $M$, respectively. Note that
$\sigma^2$ is computed at $z=0$ in this equation. We adopt $f=0.01$
following JS02. Equation (\ref{eq:fc}) suggests that the concentration
parameter becomes too small for very elongated haloes ($a/c\ll
1$). Since the fitting formula of $f_c$ was derived at $f_c\ga 0.3$
(see JS02), we modified the prefactor in equation (\ref{eq:ce}) from
$f_c$ to ${\rm Max}[f_c,0.3]$ in order to avoid unrealistically small
values of the concentration parameter.   

We need to specify the orientation of each halo relative to the
line-of-sight direction to compute lensing properties. The orientation
of dark haloes can be specified by the following two angles; $\alpha$
($0<\alpha<\pi$) defined by an angle between the major axis of the
triaxial halo and the line-of-sight direction, and $\beta$ ($0<\beta<2\pi$)
which represents a rotation angle in a plane perpendicular to the major axis
\citep[see][]{oguri03,oguri04}. Assuming that the orientation of each
halo is random, the PDFs of these angles are given by 
\begin{equation}
p(\alpha)=\frac{\sin\alpha}{2},
\end{equation}
\begin{equation}
p(\beta)=\frac{1}{2\pi},
\end{equation}
We perform the projection of the triaxial halo following the procedure 
given by \citet{oguri03} and compute the projected convergence and shear
maps for a given source redshift $z_s$:  
\begin{equation}
\kappa(x,y)=\frac{b_{\rm TNFW}}{2}f_{\rm NFW}\left(\frac{1}{R_0}
\sqrt{\frac{x^2}{q_x}+\frac{y^2}{q_y^2}}\right),
\label{eq:kappa}
\end{equation}
\begin{equation}
b_{\rm TNFW}=\frac{4\delta_{\rm cd}\rho_{\rm crit}(z)R_0}{\sqrt{f}\Sigma_{\rm cr}},
\end{equation}
\begin{equation}
f_{\rm NFW}(x)=\int_0^\infty dz \frac{\left\{1+(x^2+z^2)/x_t^2\right\}^{-2}}{\sqrt{x^2+z^2}(1+\sqrt{x^2+z^2})^2},
\end{equation}
where $\Sigma_{\rm cr}$ is the critical surface mass density for
lensing and $x_t=4r_{\rm vir}/R_0$ is the truncation radius. The
parameters $q_x$, $q_y$, and $f$ are complicated functions of the axis
ratios and the projection direction \citep[see][for explicit
  expressions]{oguri03}.  The axis ratio of the
projected mass distribution is given by $q\equiv q_y/q_x$. 
This model introduces the ellipticity in the projected density, and
therefore does not suffer from unphysical mass distributions that are
seen if the ellipticity is introduced in the lens potential
\citep[e.g.,][]{golse02}. Since
critical curves of projected triaxial haloes are in general neither
circles nor ellipses,  the definition of the Einstein radii for these
systems are not trivial. In this paper we compute the Einstein radii
as follows. First we calculate distances from the halo centre to the
(outer) critical curve along the major and minor axes of projected
two-dimensional density distribution, which we denote $\theta_x$ and
$\theta_y$, respectively. Then we estimate the Einstein radius of the
system by the geometric mean of these two distances: 
\begin{equation}
 \theta_{\rm E}=\sqrt{\theta_x\theta_y}.
\label{eq:ein}
\end{equation}
By computing Einstein radii for all the massive dark haloes we have
randomly generated, we obtain a mock all-sky catalogue of Einstein
radii. For each model we consider, we generate 300 of all-sky
realisations in order to assess the cosmic variance of the largest
Einstein radii in the universe. 

\begin{figure}
\begin{center}
 \includegraphics[width=0.9\hsize]{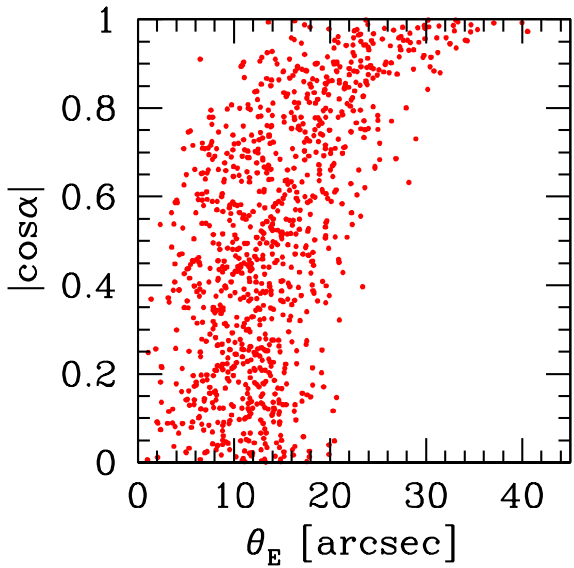}
\end{center}
\caption{The distribution of the Einstein radii $\theta_{\rm E}$ for
 1,000 triaxial haloes with fixed virial mass $M_{\rm vir}=2\times
 10^{15}M_\odot$. The haloes are located at $z_l=0.3$, and the
 source redshift is assumed to be $z_s=1$. The distribution is shown
 as a function of $|\cos\alpha|$, where $\alpha$ is an angle between
 the major axis of each halo and the line-of-sight direction. For
 comparison, the virial radius of the cluster corresponds to
 $\theta\sim640''$. 
\label{fig:tri}}
\end{figure}

To demonstrate the importance of triaxiality on this study,  we
compute Einstein radii of 1,000 massive dark haloes with the virial mass
$M_{\rm vir}=2\times 10^{15}M_\odot$ and the redshift $z_l=0.3$. The
concentration parameter, axis ratios, and the orientation with respect
to the line-of-sight direction of each halo are randomly generated using
the PDFs described above. We show the resulting distribution of
$\theta_{\rm E}$ in Figure~\ref{fig:tri}. The Figure indicates that
haloes of the same mass can have a wide range of the Einstein
radii. They are correlated with the orientation of the halo such that
largest Einstein radii are caused only when the major axis of haloes is
almost aligned with the line-of-sight direction ($|\cos\alpha|\sim
1$), implying a strong orientation bias in large lens clusters.

\section{Largest Einstein radius and properties of the lensing cluster}
\label{sec:largest}

\subsection{Probability distribution of the largest Einstein radius}

\begin{figure}
\begin{center}
 \includegraphics[width=0.95\hsize]{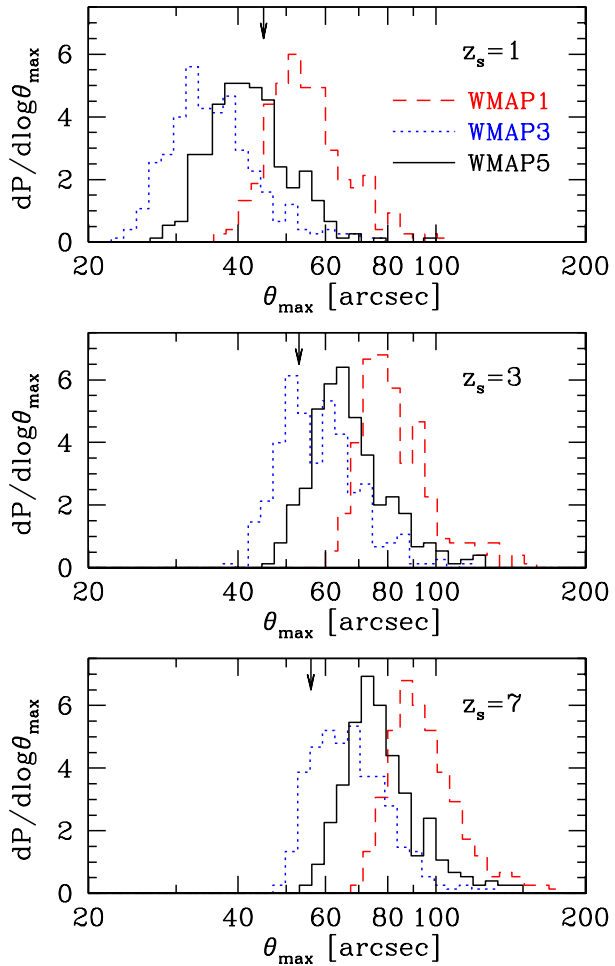}
\end{center}
\caption{Probability distributions of the largest Einstein radius
 $\theta_{\rm max}$, constructed from 300 Monte-Carlo realisations of
 all-sky massive cluster catalogues. From top to bottom panels, source
 redshifts are assumed to be $z_s=1$, $3$, and $7$. Results for three
 different cosmological models, WMAP1 ({\it dashed}), WMAP3 ({\it
 dotted}), and WMAP5 ({\it solid}) are shown. Arrows indicate the
 values of the Einstein radii of A1689, which have the largest known 
 Einstein radii. Note that they correspond to the lower limits of
 $\theta_{\rm max}$ in  observations (see text for details).   
\label{fig:hist_ein}}
\end{figure}

First we take the cluster that has the largest Einstein radius from each
all-sky realisation. From the 300 realisations for each cosmological
model, we can not only construct a probability distribution of the
largest Einstein radius in the universe, but also obtain expected
properties of the lensing cluster. In what follows, we consider three
source redshifts, $z_s=1$, $3$, and $7$. The source redshift of
$z_s=1$ is more relevant to typical giant luminous arcs or weak
lensing studies, whereas results for $z_s=7$ are more important in
searching for high-redshift galaxies near critical curves. The
redshift distribution of strongly-lensed faint background galaxies in
massive clusters has a peak at $z_s\sim 3$
\citep[e.g.,][]{broadhurst05b}. The  
probability distributions of the largest Einstein radius in all-sky,
$\theta_{\rm max}$, are shown in Figure~\ref{fig:hist_ein}. As
expected, $\theta_{\rm max}$ is quite dependent on the cosmological
model. For the source redshift $z_s=1$, the median of the largest
Einstein radius is $\theta_{\rm max}=54''$ for WMAP1, $\theta_{\rm
  max}=35''$ for WMAP3, and $\theta_{\rm max}=42''$ for WMAP5. 
The different values of $\sigma_8$ ($\sigma_8=0.9$ for WMAP1,
$\sigma_8=0.76$ for WMAP3, and $\sigma_8=0.8$ for WMAP5) change the 
abundance of massive dark haloes and its redshift evolution
drastically, which is why the largest Einstein radius is sensitive to
$\sigma_8$. It is also found that the value of
the largest Einstein radius is quite dependent on the source redshift
as well: $\theta_{\rm max}$ for $z_s=7$ is approximately twice as
large as that for $z_s=1$.   

\begin{figure}
\begin{center}
 \includegraphics[width=0.95\hsize]{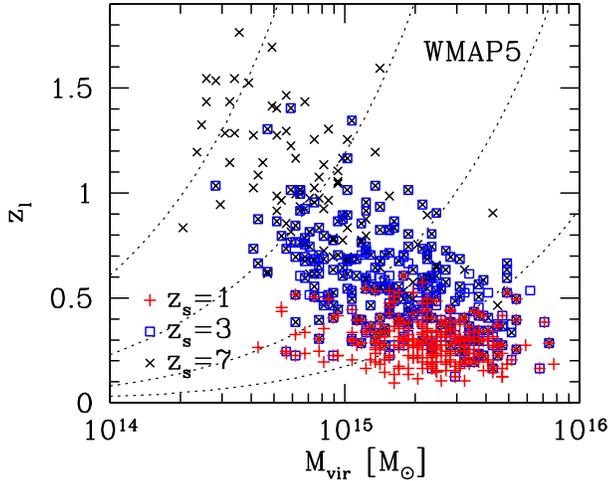}
\end{center}
\caption{The distribution of the mass and redshift of the cluster
  producing the largest Einstein radius. Each point corresponds to one
  Monte-Carlo realisation. Results are shown for three different
  redshifts, $z_s=1$ ({\it pluses}), $3$ ({\it open squares}), $7$
  ({\it crosses}). The WMAP5 cosmology is assumed in this plot.
  Also plotted are contours of constant X-ray fluxes, inferred
  from the correlation between bolometric X-ray luminosities and
  halo virial masses \citep[][assuming no redshift
  evolution]{shimizu03}. From right to left, the contours indicate 
  X-ray fluxes of $f_X=10^{-11}$, $10^{-12}$, $10^{-13}$, and
  $10^{-14}$~${\rm erg\,s^{-1}cm^{-2}}$.  
\label{fig:massz}}
\end{figure}

\begin{figure}
\begin{center}
 \includegraphics[width=0.95\hsize]{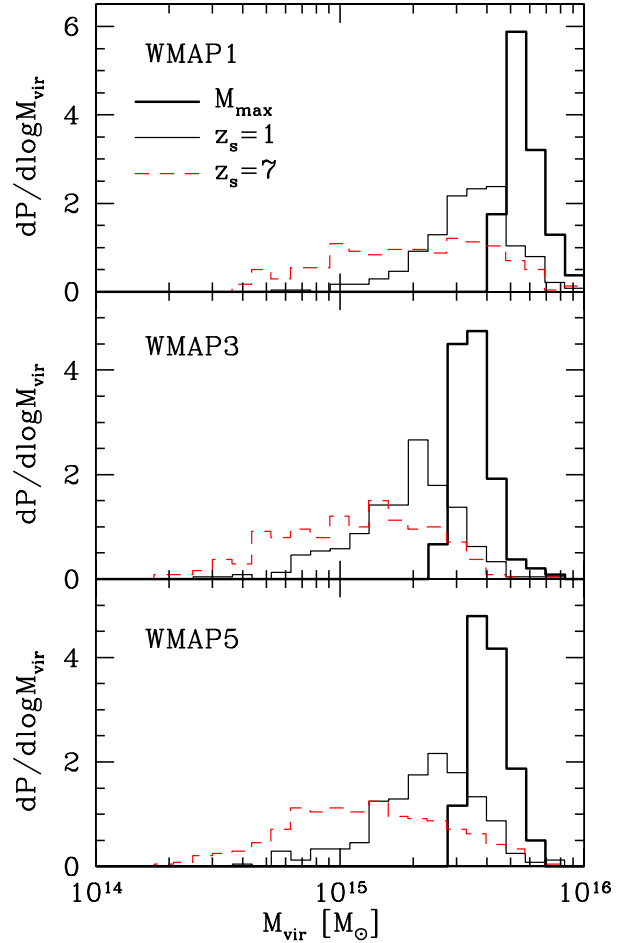}
\end{center}
\caption{Probability distributions of masses $M_{\rm vir}$ of the
  most massive cluster in the universe ({\it thick solid}) and the
  cluster producing the largest Einstein radius ({\it thin solid} for 
  $z_s=1$ and {\it dashed} for $z_s=7$). From top to bottom panels, 
  we show results for the WMAP1, WMAP3, and WMAP5 cosmologies,
  respectively. 
\label{fig:hist_mmax}}
\end{figure}

The cluster which has the largest known Einstein radius to date is
A1689. It is a massive cluster located at low redshift ($z_l=0.18$),
and its Einstein radius is well constrained from many multiply-lensed
background galaxies and weak lensing to be $\theta_{\rm E}=45''$ for
$z_s=1$ \citep{tyson90,miralda95,clowe01,broadhurst05a,broadhurst05b,umetsu08}.
From the best-fit mass model of \citet{umetsu08}, we derive $\theta_{\rm
  E}=53''$ for $z_s=3$ and $\theta_{\rm E}=56''$ for $z_s=7$. We
emphasise that these values should be viewed as the lower limits of
$\theta_{\rm max}$, since even larger lens clusters may be
discovered in the future. Nevertheless, we compare these values with
our theoretical predictions in Figure~\ref{fig:hist_ein}. For $z_s=1$,
the Einstein radius of A1689 is quite consistent with our prediction
for our fiducial cosmological model, WMAP5. On the other hand it is
slightly larger (smaller) than our prediction for  WMAP3 (WMAP1), but
is within 90 percentile for both WMAP1 and WMAP3. In contrast, the
observed values are consistent with WMAP3 and lower than WMAP1 and
WMAP5 for $z_s=3$. For the higher source redshift $z_s=7$, the
Einstein radius of A1689 is even smaller than our prediction for
WMAP3. An implication of this is that there may exist a cluster with
the Einstein radius larger than that of A1689 for that source
redshift. We note that cosmological models which predict $\theta_{\rm
  max}$ close to the Einstein radius of A1689 for $z_s=7$, such as
models with $\sigma_8$ even smaller than WMAP3, should predict too
small $\theta_{\rm max}$ for $z_s=1$ to be consistent with A1689. 

Whilst the better match to WMAP3/WMAP5 is consistent with recent
results from the cluster abundance
\citep[e.g.,][]{dahle06,gladders07,mantz08}, the large discrepancy
between the predicted Einstein radii for WMAP1 and those of A1689 does
not necessarily exclude WMAP1 cosmology as the current observed
$\theta_{\rm max}$ correspond to the lower limits. Complete surveys of
large lens clusters are necessary to extract useful cosmological
information from this statistics (see \S\ref{sec:survey}).

\subsection{Cluster mass and redshift}
\label{sec:clumz}

Next we examine the expected mass and redshift distribution of the
cluster which produces the largest Einstein radii. The distributions
shown in Figure~\ref{fig:massz} indicate that wide ranges of mass and
redshift are possible. In particular, it is worth noting that the mass
of the cluster can be as small as $M_{\rm vir}<10^{15}M_\odot$. In
addition, the cluster can be located at quite high-redshifts, up to
$z_l\sim 1$ and beyond, in the case of $z_s=3$ and $7$, which will be
difficult to access via X-ray observations with currently operating
telescopes (see also discussions in \S\ref{sec:survey}).
On the other hand, for $z_s=1$  the lens cluster is likely
to be located at $z_l<0.5$.  Clearly the diversity of the mass and
redshift is a consequence of halo triaxiality, which we will explore
later.  

Here a natural question to ask is whether or not the cluster having
the largest Einstein radius is the most massive cluster in the
universe. To check this we take the most massive cluster in each
realisation and construct the probability distribution of its mass. It
is then compared with the PDF of the mass of the largest lens
cluster. We show the result in Figure~\ref{fig:hist_mmax}. As we
discussed, the cluster with $\theta_{\rm max}$ has a wider range of
the mass and thus is not necessarily the most massive cluster in the
universe. However, the overlap of the PDFs at the high-mass end for
all the three cosmological models suggests that in some cases the
largest lens corresponds to the most  massive cluster. It is
interesting to note that the PDFs for $z_s=7$ extend to lower masses
than those for $z_s=1$. This is because the most massive clusters are
typically located at $z_l\sim 0.1-0.4$, whereas the geometrical
lensing efficiency for the source redshift $z_s=7$ is the highest at
around $z_l\sim 1$ where clusters are on average less massive (see
also Figure~\ref{fig:massz}). 

\subsection{Expected properties of the lensing cluster}
\label{sec:bias}

\begin{figure*}
\begin{center}
 \includegraphics[width=0.9\hsize]{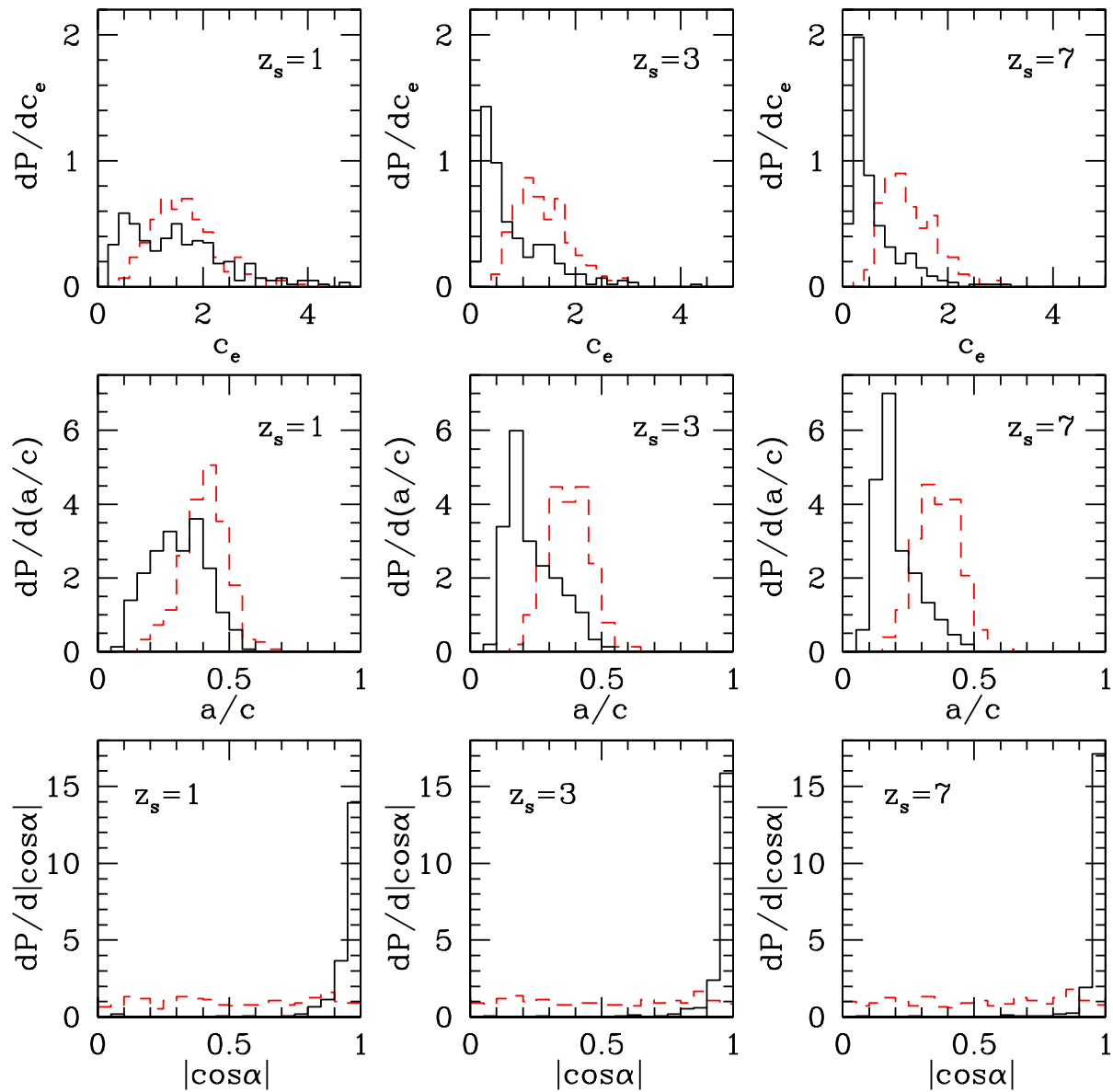}
\end{center}
\caption{Probability distributions of the triaxial concentration
  parameter $c_e$ ({\it top}), the triaxial axis ratio $a/c$ ({\it
  centre}), and the angle between the  major axis and the
  line-of-sight direction $\alpha$ ({\it bottom}) for the cluster
  producing the largest Einstein radius are plotted. For comparison, 
  the PDFs for typical (unbiased) clusters with the same mass and
  redshift distributions are shown by dashed histograms. From left to
  right panels, we change the source redshift from $1$ to
  $7$. Although results shown here are for the WMAP5 cosmology, we
  confirmed that the PDFs for the other two cosmologies are similar.  
\label{fig:hist_par_w5}}
\end{figure*}

\begin{figure*}
\begin{center}
 \includegraphics[width=0.9\hsize]{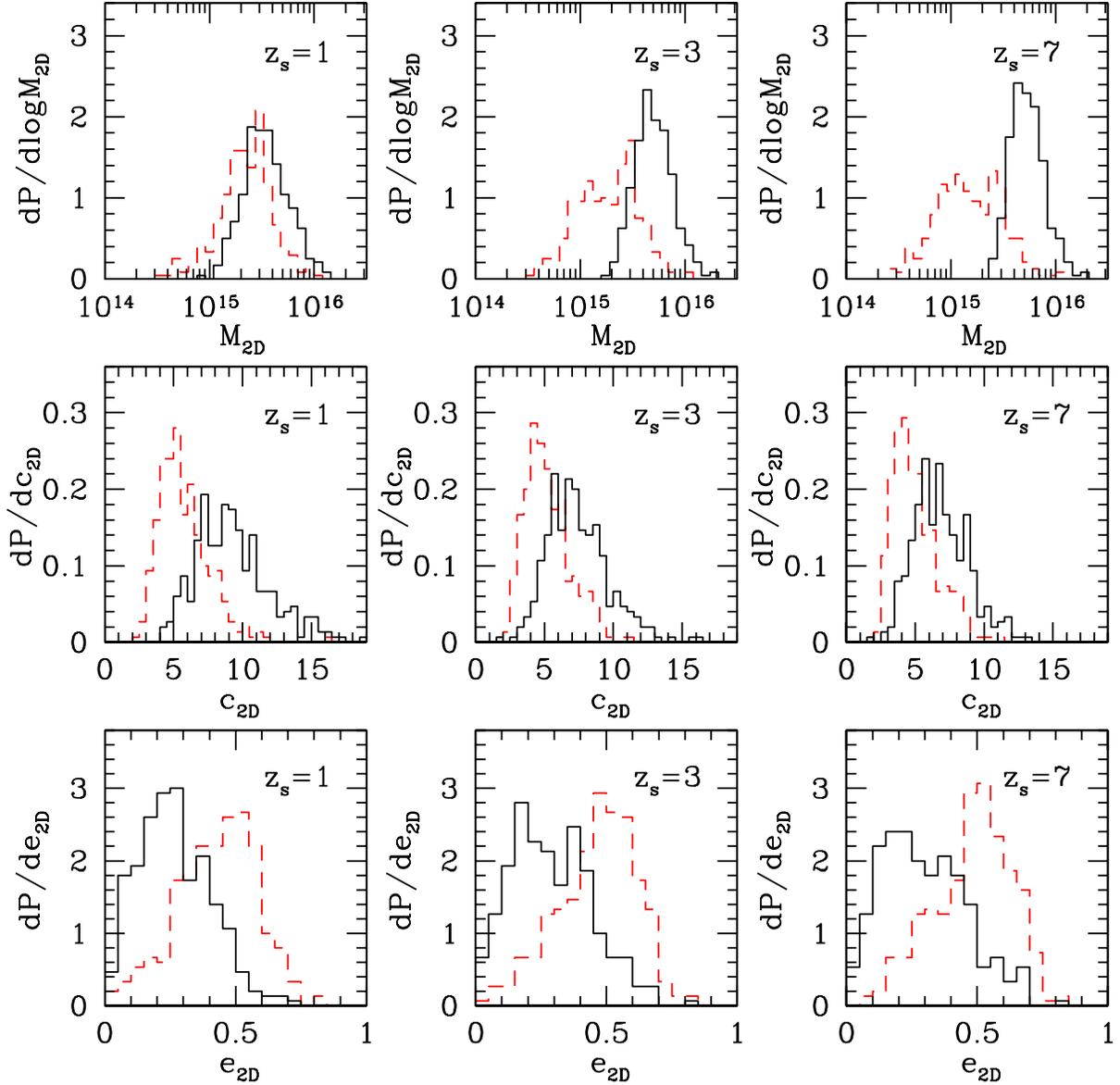}
\end{center}
\caption{Similar to Figure~\ref{fig:hist_par_w5}, but the PDFs of the
 projected two-dimensional virial mass $M_{\rm 2D}$, the projected 
two-dimensional spherical concentration parameter $c_{\rm 2D}$, and
the projected ellipticity $e_{\rm 2D}$ are shown. 
\label{fig:hist_twod_w5}}
\end{figure*}

\begin{table*}
 \caption{The largest Einstein radii and expected properties of
 the lensing clusters. We show median values and 68\% confidence
 intervals estimated from 300 Monte-Carlo realisations. Values in
 parentheses indicate corresponding parameter values for typical
 clusters: they are estimated by adopting same mass and redshift
 distributions as those of the lens cluster and re-assigning
 concentrations, axis ratios, and orientations according to the
 PDFs.\label{table:ein}} 
 \begin{tabular}{@{}ccccccccccc}
  \hline
   Model & $z_s$ & $\theta_{\rm max}$ &  $M_{\rm vir}$ 
   & $z_l$ & $c_e$ & $a/c$ & $|\cos\alpha|$ 
   & $M_{\rm 2D}$ & $c_{\rm 2D}$ & $e_{\rm 2D}$ \\
   & & [arcsec] & [$10^{15}M_\odot$] &  & & & & [$10^{15}M_\odot$] & & \\
   \hline
  WMAP1 & 1 & $54^{+12}_{-7}$ & $3.57^{+1.36}_{-1.32}$ & $0.28^{+0.11}_{-0.09}$ & $1.57^{+0.98}_{-0.88}$ & $0.32^{+0.10}_{-0.10}$ & $0.97^{+0.02}_{-0.09}$ & $4.58^{+3.03}_{-1.49}$ & $8.91^{+2.64}_{-2.11}$ & $0.26^{+0.18}_{-0.12}$\\ 
& & & & &($1.63^{+0.65}_{-0.58}$) & ($0.41^{+0.09}_{-0.09}$) & ($0.51^{+0.34}_{-0.32}$) & ($3.31^{+1.91}_{-1.24}$) & ($5.59^{+2.12}_{-1.53}$) & ($0.43^{+0.16}_{-0.17}$) \\
  & 3 & $80^{+15}_{-9}$ & $2.83^{+1.87}_{-1.54}$ & $0.47^{+0.25}_{-0.17}$ & $0.66^{+0.87}_{-0.35}$ & $0.23^{+0.11}_{-0.07}$ & $0.98^{+0.02}_{-0.06}$ & $6.55^{+3.71}_{-1.97}$ & $7.03^{+2.27}_{-1.93}$ & $0.30^{+0.16}_{-0.15}$ \\
& & & & & ($1.42^{+0.65}_{-0.46}$) & ($0.39^{+0.11}_{-0.09}$) & ($0.48^{+0.34}_{-0.34}$) & ($2.84^{+1.86}_{-1.46}$) & ($5.00^{+2.25}_{-1.22}$) & ($0.46^{+0.14}_{-0.18}$) \\
 & 7 & $92^{+18}_{-10}$ & $2.05^{+2.23}_{-1.16}$ & $0.62^{+0.35}_{-0.24}$ & $0.44^{+0.67}_{-0.20}$ & $0.19^{+0.12}_{-0.05}$ & $0.98^{+0.02}_{-0.05}$ & $7.06^{+3.39}_{-1.77}$ & $6.36^{+2.25}_{-1.53}$ & $0.33^{+0.16}_{-0.16}$ \\
& & & & & ($1.29^{+0.60}_{-0.42}$) & ($0.39^{+0.08}_{-0.09}$) & ($0.48^{+0.38}_{-0.35}$) & ($2.26^{+2.37}_{-1.26}$) & ($4.88^{+1.66}_{-1.38}$) & ($0.48^{+0.14}_{-0.19}$) \\
 WMAP3 & 1 & $35^{+8}_{-6}$ & $2.05^{+0.91}_{-0.92}$ & $0.29^{+0.13}_{-0.09}$ & $1.23^{+1.19}_{-0.76}$ & $0.29^{+0.12}_{-0.11}$ & $0.98^{+0.02}_{-0.08}$ & $2.93^{+2.32}_{-1.07}$ & $8.67^{+2.96}_{-2.35}$ & $0.26^{+0.18}_{-0.12}$ \\
& & & & & ($1.52^{+0.64}_{-0.50}$) & ($0.40^{+0.09}_{-0.08}$) & ($0.54^{+0.29}_{-0.35}$) & ($1.95^{+1.29}_{-0.93}$) & ($5.37^{+1.91}_{-1.35}$) & ($0.45^{+0.15}_{-0.15}$) \\
  & 3 & $57^{+13}_{-8}$ & $1.48^{+1.09}_{-0.77}$ & $0.53^{+0.26}_{-0.19}$ & $0.57^{+0.66}_{-0.27}$ & $0.22^{+0.10}_{-0.07}$ & $0.98^{+0.01}_{-0.04}$ & $4.18^{+2.05}_{-1.29}$ & $6.92^{+2.08}_{-1.60}$ & $0.28^{+0.16}_{-0.13}$ \\
& & & & & ($1.31^{+0.57}_{-0.46}$) & ($0.38^{+0.09}_{-0.09}$) & ($0.48^{+0.35}_{-0.33}$) & ($1.59^{+1.25}_{-0.78}$) & ($4.76^{+1.97}_{-1.18}$) & ($0.49^{+0.14}_{-0.21}$) \\
  & 7 & $67^{+13}_{-10}$ & $1.23^{+1.23}_{-0.69}$ & $0.62^{+0.32}_{-0.23}$ & $0.44^{+0.60}_{-0.20}$ & $0.19^{+0.11}_{-0.06}$ & $0.99^{+0.01}_{-0.04}$ & $4.47^{+1.78}_{-1.37}$ & $6.65^{+1.94}_{-1.67}$ & $0.28^{+0.17}_{-0.13}$ \\
& & & & & ($1.27^{+0.54}_{-0.43}$) & ($0.36^{+0.09}_{-0.07}$) &
   ($0.48^{+0.34}_{-0.31}$) & ($1.35^{+1.32}_{-0.73}$) &
   ($4.85^{+1.67}_{-1.30}$) & ($0.50^{+0.13}_{-0.17}$) \\
WMAP5 & 1 & $42^{+9}_{-7}$ & $2.35^{+1.21}_{-0.93}$ & $0.29^{+0.10}_{-0.10}$ & $1.43^{+0.98}_{-0.87}$ & $0.31^{+0.11}_{-0.12}$ & $0.98^{+0.02}_{-0.06}$ & $3.35^{+2.55}_{-1.17}$ & $8.99^{+2.78}_{-2.28}$ & $0.26^{+0.15}_{-0.13}$ \\
& & & & & ($1.61^{+0.62}_{-0.55}$) & ($0.41^{+0.08}_{-0.09}$) & ($0.50^{+0.36}_{-0.33}$) & ($2.29^{+1.45}_{-0.98}$) & ($5.42^{+2.19}_{-1.31}$) & ($0.45^{+0.14}_{-0.17}$) \\
 & 3 & $65^{+15}_{-8}$ & $1.78^{+1.61}_{-0.97}$ & $0.52^{+0.23}_{-0.20}$ & $0.58^{+0.90}_{-0.29}$ & $0.21^{+0.14}_{-0.06}$ & $0.98^{+0.01}_{-0.04}$ & $4.87^{+2.80}_{-1.55}$ & $7.12^{+2.34}_{-1.80}$ & $0.28^{+0.17}_{-0.15}$ \\
& & & & & ($1.31^{+0.56}_{-0.45}$) & ($0.37^{+0.08}_{-0.08}$) & ($0.50^{+0.36}_{-0.34}$) & ($1.94^{+1.55}_{-1.05}$) & ($4.97^{+2.03}_{-1.30}$) & ($0.49^{+0.12}_{-0.19}$) \\
 & 7 & $76^{+17}_{-9}$ & $1.29^{+1.53}_{-0.67}$ & $0.66^{+0.36}_{-0.26}$ & $0.41^{+0.66}_{-0.18}$ & $0.18^{+0.12}_{-0.04}$ & $0.99^{+0.01}_{-0.03}$ & $5.15^{+2.65}_{-1.43}$ & $6.64^{+2.25}_{-1.73}$ & $0.28^{+0.19}_{-0.15}$ \\
& & & & & ($1.18^{+0.59}_{-0.38}$) & ($0.36^{+0.08}_{-0.08}$) & ($0.53^{+0.34}_{-0.36}$) & ($1.45^{+1.53}_{-0.71}$) & ($4.71^{+1.88}_{-1.25}$) & ($0.50^{+0.14}_{-0.19}$) \\
 \hline
 \end{tabular}
\end{table*}

It has been argued that the population of lenses is markedly different
from that of nonlenses in several ways  
\citep[e.g.,][]{oguri05b,hennawi07b,moeller07,fedeli07b,rozo08b}. 
The largest Einstein radius represents the most extreme case of
lensing clusters, which suggests that the cluster population may be
biased even more strongly. Here we quantify the lensing bias in the
cluster producing $\theta_{\rm max}$ from our Monte-Carlo
realisations.   

As discussed above, an important parameter here is the orientation of
the cluster, specifically the angle $\alpha$ between the major axis
and the line-of-sight direction. Another important parameter is the
minor-to-major axis ratio, $a/c$, since the projection effect is
stronger for more triaxial clusters. Finally the concentration
parameter of the triaxial model, $c_e$, should also be of interest
because the strong lensing efficiency is known to be sensitive to the
halo concentration.  

Figure~\ref{fig:hist_par_w5} shows probability distributions of
these three parameters for the cluster having $\theta_{\rm max}$.
We also show the PDFs for a ``typical'' cluster which has the same
mass and redshift probability distributions as those of the lens 
cluster but the axis ratio, the concentration, and the orientation are
re-assigned from their original PDFs. Thus the comparison of the lens
and typical cluster PDFs provides the degree of lensing biases. 
Strikingly, we find that the cluster is almost always aligned with the
line-of-sight direction. For instance, the probability of having
$|\cos\alpha|>0.9$ ($\alpha<25.8^\circ$) is 0.88 for $z_s=1$ and 0.95
for $z_s=7$ for the WMAP5 cosmology. It is also found that the lens
cluster is more triaxial than typical clusters. Because of the
correlation between $c_e$ and $a/c$, the triaxial concentration
parameter for the lens cluster becomes smaller. The strong biases in
the orientation and triaxiality indicate that such largest lens
cluster is indeed very unusual in terms of its internal structure and
configuration.   

The projection effect of the triaxial halo has a large impact on the 
apparent mass profile constrained from lensing data \citep{oguri05a}.
To investigate this, we characterise the projected two-dimensional (2D)
mass distribution of each triaxial cluster by the following three
parameters: the mass $M_{\rm 2D}$ and concentration parameter $c_{\rm 
2D}$ of the NFW profile, which are obtained from the projected surface
mass density by ignoring the elongation along the line-of-sight, and
the ellipticity of the surface mass density $e_{\rm 2D}$. Our
parameter $c_{\rm 2D}$ corresponds to the standard concentration
parameter which has been studied from analysis of observed lensing
clusters, and thus is useful for discussing possible high
concentrations from lens mass reconstructions
\citep[e.g.,][]{broadhurst05a}. We can relate these parameters to
those of the triaxial halo by simply comparing the expressions of the
projected surface mass densities (K. Takahashi et al., in preparation): 
\begin{equation}
 1-e_{\rm 2D}=q,
\end{equation}
\begin{equation}
 \frac{r_{\rm vir}(M_{\rm 2D})}{\sqrt{q}c_{\rm 2D}}=R_0q_x,
\end{equation}
\begin{equation}
 b_{\rm NFW}(M_{\rm 2D},c_{\rm 2D})=b_{\rm TNFW}(M_{\rm vir}, c_e),
\end{equation}
where $b_{\rm NFW}$ is the lensing strength parameter for the
spherical NFW profile. 

We show probability distributions of these 2D parameters, for both the 
cluster having $\theta_{\rm max}$ and corresponding typical
cluster, in Figure~\ref{fig:hist_twod_w5}. It is clear that the
largest lens cluster is highly biased in terms of the 2D parameters as 
well. In lensing observations the lens cluster looks more massive and
more centrally concentrated than typical clusters. Indeed this is
expected from the strong orientation bias (see above), because both
the mass and concentration of the cluster projected along the major
axis are known to be significantly overestimated
\citep{oguri05a,corless08}.  In 
addition we find that the cluster should appear rounder. Its expected
median ellipticity of  $0.25-0.3$ is significantly smaller than that of 
typical triaxial cluster, $0.45-0.5$. Again, this is because of the
orientation bias: the projected mass distribution of a triaxial halo
is most elliptical when it is projected along the middle axis, and
least elliptical when projected along major or minor axis. This means 
that the circularity of projected density distribution do not
necessarily imply the sphericity of the cluster. We summarise our
numerical results in Table~\ref{table:ein}.  

We are now in a position of discussing the high concentration
parameters observed in some lens-rich clusters. For instance, from the
strong and weak lensing data \citet{umetsu08} refined the
concentration parameter of A1689, which has the largest known Einstein
radius, to be $c_{\rm 2D}=12.7$ assuming a spherical NFW
profile.\footnote{Although \citet{limousin07} obtained somewhat
  smaller value of concentration, $c_{\rm 2D}=9.6$, form their strong
  and weak lensing analysis, it has been argued that their best-fit 
  model predicts the Einstein radius smaller than what is observed
  \citep[see][]{umetsu08}. Since the Einstein radius is of central
  interest, in this paper we adopt the result of \citet{umetsu08} as
  a fiducial best-fit model of A1689.} 
 Adopting the source redshift $z_s=1$, our model predicts that 
such a cluster should have the concentration parameter in the range
$6.7<c_{\rm 2D}<11.8$  at 68\% confidence and $5.5<c_{\rm 2D}<14.8$ at
90\% confidence. Thus we conclude that the high concentration of A1689
is consistent with the theoretical expectation based on the
F$\Lambda$CDM model at $1.2\sigma$ level. We note that the redshift of
the cluster is also consistent at $\sim 1\sigma$ level.
However the halo concentration of A1689 is still slightly larger than
the theoretical expectation. Therefore statistical studies of
concentrations for several large lens clusters, as attempted in
\citet{broadhurst08a} and \citet{broadhurst08b}, will be essential to
assess whether or not the large concentration problem is indeed existent.

\begin{figure}
\begin{center}
 \includegraphics[width=0.95\hsize]{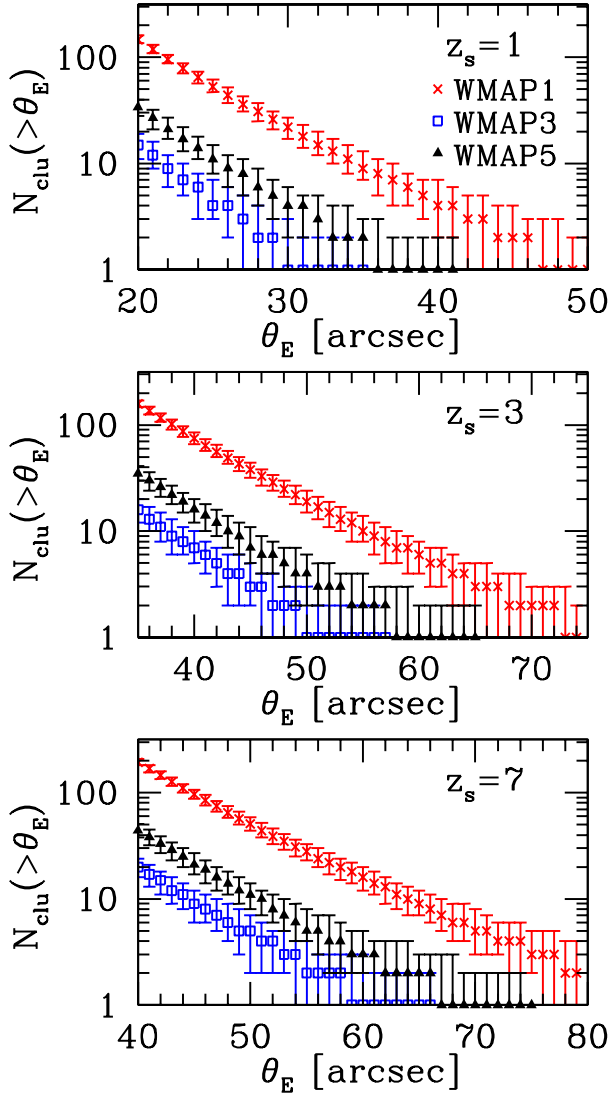}
\end{center}
\caption{Expected all-sky cumulative number distributions of large
 Einstein radii. The median and $1\sigma$ error-bars of the numbers
 are estimated from 300 Monte-Carlo realisations. Results are
 presented for all the three cosmological models, WMAP1 ({\it
 crosses}), WMAP3 ({\it squares}), and WMAP5 ({\it triangles}). From
 top to bottom panels, results for $z_s=1$, $3$, and $7$ are shown.
\label{fig:sep}}
\end{figure}

\section{Distribution of Einstein radii}
\label{sec:num}

Thus far we focused our attention on the largest Einstein radius on
the sky. Another interesting quantity to investigate is the number of
clusters which have relatively large Einstein radii. Here we derive
the expected number distribution of such large Einstein radii from our
Monte-Carlo realisations.   

In Figure~\ref{fig:sep}, we plot the cumulative number distributions for
all the three cosmological models. It is found that the number
decreases exponentially with increasing Einstein radius. As in the
case of the largest Einstein radius, the abundance of large lens
clusters is quite sensitive to the cosmological model. For instance,
the all-sky numbers of clusters which have $\theta_{\rm E}>20''$
(for $z_s=1$) are predicted to be $\sim 150$, $\sim 15$, $\sim 35$,
for WMAP1, WMAP3, and WMAP5 cosmologies, respectively. The large
difference of the cumulative numbers between WMAP1 and WMAP3 is
broadly consistent with \citet{li06,li07} who investigated strong
lensing probabilities in two different cosmological simulations. The
result suggests that it provides useful constraints on cosmological
parameters.   

\begin{figure*}
\begin{center}
 \includegraphics[width=0.9\hsize]{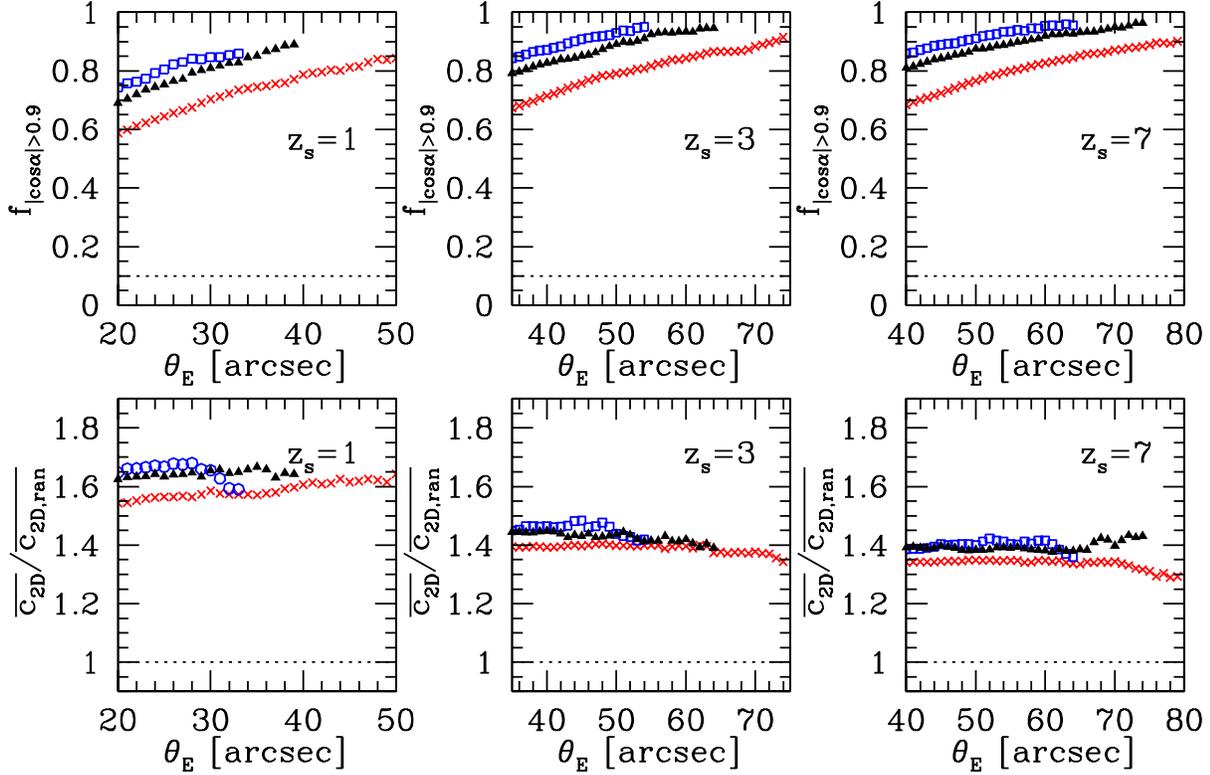}
\end{center}
\caption{The dependence of lensing biases on the limiting Einstein
radius $\theta_{\rm E}$. Each point shows lensing biases of the 
cluster orientation ({\it upper}), $f_{|\cos\alpha>0.9|}$, or that of
 the concentration parameter inferred from the projected 2D mass
 distribution ({\it lower}), $\overline{c_{\rm 2D}}/\overline{c_{\rm
 2D,ran}}$, in clusters having the Einstein radii larger than
 $\theta_{\rm E}$. See text for the definitions of these
 parameters. Horizontal dotted lines indicate values in the case with no
 bias. Crosses, open squares, and filled triangles denote results for
 WMAP1, WMAP3, and WMAP5 models, respectively.  We consider the source
 redshift of $z_s=1$ ({\it left}), $3$ ({\it centre}), and $7$ 
 ({\it right}). \label{fig:bias_sep}}
\end{figure*}

One of the main findings about the largest Einstein radius was that
the lens clusters constitute a highly biased population (see
Section~\ref{sec:bias}). One might expect that the strong bias is due to
the rareness of such largest lens, and thus it is interesting to check
lensing biases for more common lens events. In
Figure~\ref{fig:bias_sep}, we show biases in clusters which have
Einstein radii larger than certain values. Here we adopt two 
parameters to quantify lensing biases. One is a measure of the
orientation bias, $f_{|\cos\alpha>0.9|}$, which is defined by the
fraction that the angle between the major axis of the cluster and the
line-of-sight direction satisfies $|\cos\alpha|>0.9$. If the
orientation of clusters is completely random, we should have
$f_{|\cos\alpha>0.9|}\approx 0.1$. The other parameter is to describe
the 2D concentration bias, $\overline{c_{\rm 2D}}/\overline{c_{\rm
    2D,ran}}$, which is the ratio of median 2D concentration
parameters (see also Section~\ref{sec:bias}) among large lens clusters
and corresponding typical clusters with similar mass and redshift
distributions. The parameter becomes unity if no lensing bias is
present.    

The result shown in Figure~\ref{fig:bias_sep} indicates that clusters 
with large Einstein radii are similarly highly biased as the cluster
producing the largest Einstein radii in all-sky. We find that the degree
of the orientation bias depends on the limiting Einstein radius. 
Populations of clusters with smaller limiting Einstein radius, which are
regarded to represent less extreme populations of lensing clusters, are
less biased in terms of their orientations. On the other hand, the
degree of the 2D concentration bias does not depend strongly on the
limiting Einstein radius. The bias in the 2D concentration parameter
comes both from the enhancement of the apparent concentration due to the
orientation bias and from the bias in the 3D triaxial concentration
parameter. The enhancement of $c_{\rm 2D}$ due to the orientation bias
is larger for more triaxial haloes, however because of the
anti-correlation between the axis ratio and concentration
\citep{jing02} such haloes are intrinsically less concentrated (i.e.,
have smaller $c_e$). The behavior is therefore expected to reflect the
complicated combination of these two biases which are more or less
counteract with each other. 

Our results can be compared with those of \citet{hennawi07b} who
analysed lensing biases using ray-tracing of $N$-body simulations. 
Their qualitative results are similar to ours: they found that lensing
clusters tend to have the major axis aligned with the line-of-sight and
larger 2D concentrations. However, the quantitative results are
different. Their orientation bias of $f_{|\cos\alpha>0.9|}\sim 0.25$ and
2D concentration bias of $\overline{c_{\rm 2D}}/\overline{c_{\rm
2D,ran}}\sim 1.34$ are smaller than our results (see
Figure~\ref{fig:bias_sep}). We ascribe this difference to the different
definitions of the lens cluster populations. \citet{hennawi07b} derived
the distributions for lens clusters by calculating those from all
clusters with a weight of lensing cross sections, without any
restriction to the Einstein radii. Therefore their results are relevant
to more common lens clusters with smaller Einstein radii, say
$10''-15''$, whereas our results are applicable only to superlens
clusters with unusually large Einstein radii. Our finding that the
orientation bias decreases with decreasing Einstein radius is
consistent with this interpretation.    
  
\section{Primordial Non-Gaussianity}
\label{sec:ng}

The results presented so far are based on standard universes evolved
from Gaussian initial conditions. Since the abundance of massive
clusters and its redshift evolution are known to be very sensitive to
primordial non-Gaussianities
\citep[e.g.,][]{matarrese00,verde01,mathis04,grossi07,sadeh07,fedeli08,dalal08}, 
our statistics are also expected to be dependent on primordial
non-Gaussianities. The effect of primordial non-Gaussianities is
particularly of importance given possible detections in the cosmic 
microwave background (CMB) anisotropies \citep[e.g.,][]{vielva04,yadav08}.
In this section, we repeat the same calculations conducted in the
previous sections, but including levels of primordial non-Gaussianities 
currently of interest.    

In order to quantify the effect of primordial non-Gaussianities, in this
paper we adopt the local non-Gaussianity of the following form
\citep[e.g.,][]{komatsu01,bartolo04}:
\begin{equation}
 \Phi=\phi+f_{\rm NL}\left(\phi^2-\langle\phi^2\rangle\right),
\end{equation}
where $\Phi$ is the curvature perturbation and $\phi$ is an auxiliary 
random-Gaussian field. The level of primordial non-Gaussianity is
characterised by $f_{\rm NL}$, which we assume constant. In this model
positive $f_{\rm NL}$ corresponds to positive skewness of the density
field. In observation, CMB anisotropies 
have constrained its value to be $|f_{\rm NL}| \la \mathcal{O}(100)$. For
instance, \citet{komatsu03} derived $-58<f_{\rm NL}<134$ at 95\%
confidence from the temperature map of the WMAP first-year data. More
recently, \citet{yadav08} claimed the detection of positive $f_{\rm
NL}$, $26.9<f_{\rm NL}<146.7$ at 95\% confidence, from the improved
analysis of the WMAP three-year data. On the other hand, 
\citet{komatsu08} claimed that the WMAP five-year data are marginally
consistent with Gaussian fluctuations at 95\% confidence, $-9<f_{\rm
  NL}<111$. 

We follow \citet{dalal08} to calculate the number of massive haloes in
this non-Gaussian model. From analytic considerations and $N$-body
simulations, they derived a simple fitting formula of the mass function:
\begin{equation}
 \frac{dn_{\rm NG}}{dM}=\int \frac{dM_0}{M_0}\frac{dn}{dM_0}
\frac{1}{\sqrt{2\pi}\sigma_f}\exp\left[-\frac{(M/M_0-f_M)^2}
{2\sigma_f^2}\right],
\end{equation}
where $dn/dM_0$ is a halo mass function with a Gaussian initial
condition, which we adopt equation (\ref{eq:mf}), and 
\begin{equation}
 f_M=1+1.3\times 10^{-4}f_{\rm NL}\sigma_8\sigma_{M_0}^{-2},
\end{equation}
\begin{equation}
 \sigma_f=1.4\times 10^{-4}(|f_{\rm NL}|\sigma_8)^{0.8}\sigma_{M_0}^{-1}.
\end{equation}
From this expression, we can see that the positive $f_{\rm NL}$
results in the enhancement of the abundance of massive clusters.

\begin{figure}
\begin{center}
 \includegraphics[width=0.95\hsize]{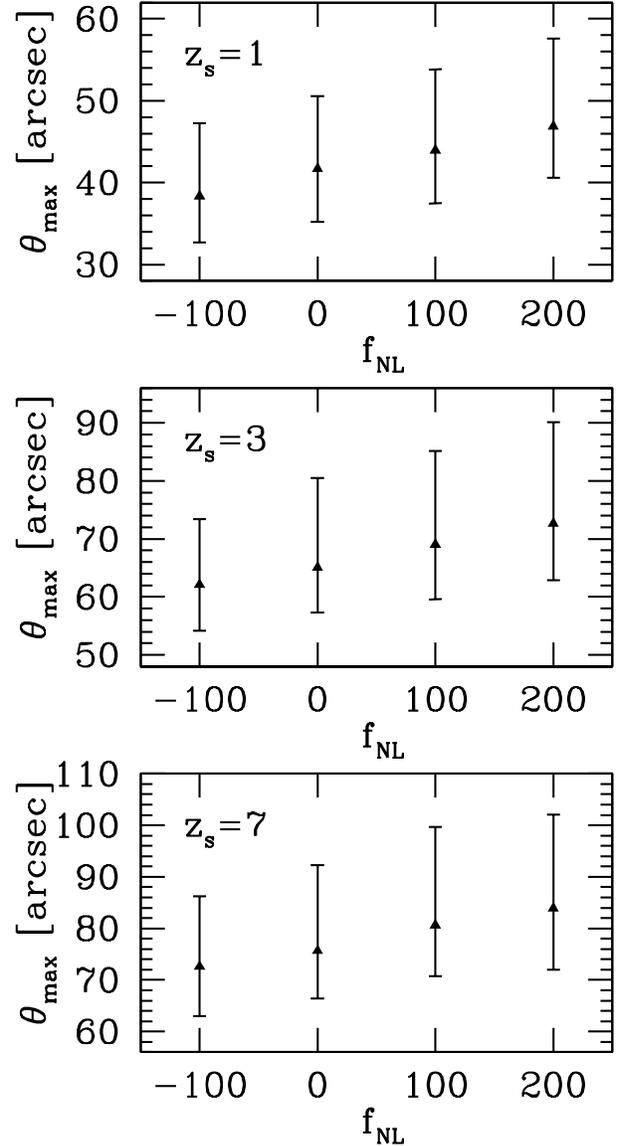}
\end{center}
\caption{The largest Einstein radius in all-sky as a function of
  primordial non-Gaussianities $f_{\rm NL}$. WMAP5 cosmology is
  assumed. The error-bars show 68\% confidence estimated from 300
 Monte-Carlo realisations. The source redshifts are assumed  to be
 $z_s=1$, $3$, and $7$ from top to bottom panels. 
\label{fig:ein_ng}}
\end{figure}

\begin{figure}
\begin{center}
 \includegraphics[width=0.95\hsize]{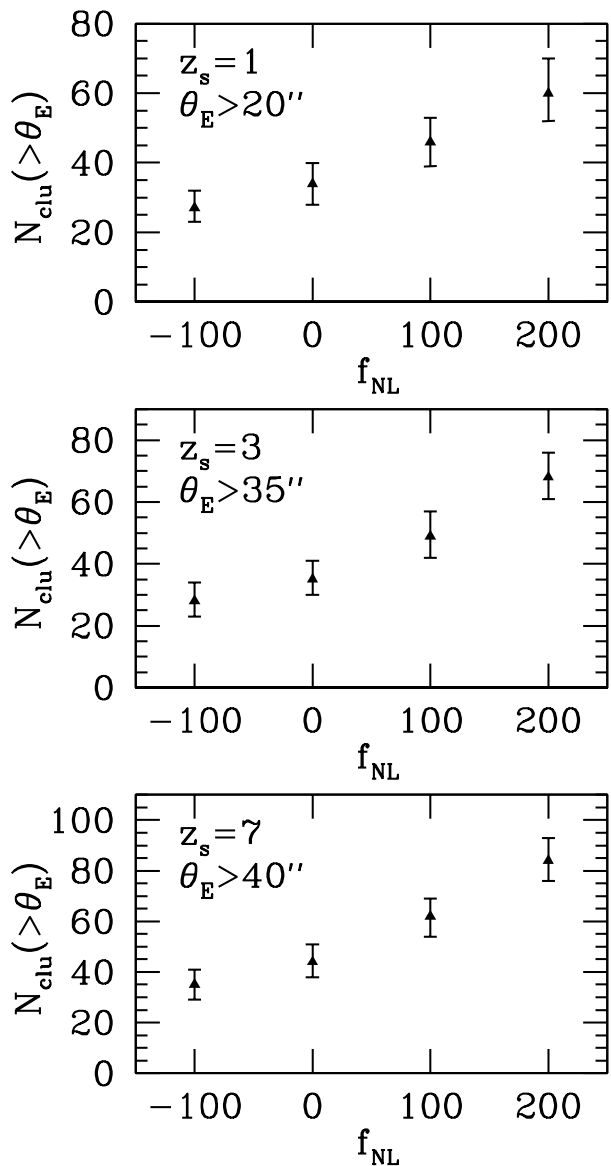}
\end{center}
\caption{Similar to Figure~\ref{fig:ein_ng}, but the numbers of
  clusters with the Einstein radii larger than certain values are
  plotted. The limiting Einstein radii are $20''$ ($z_s=1$), $35''$
  ($z_s=3$), and $40''$ ($z_s=7$).
\label{fig:sep_ng}}
\end{figure}

Primordial non-Gaussianities affect not only the abundance of massive
clusters but also their formation histories. Since the concentrations of
dark haloes are correlated with their mass assembly histories
\citep{wechsler02}, primordial non-Gaussianities should have an impact
on the halo concentration parameter as well. Indeed, $N$-body
simulations done by \citet{avila-reese03} showed that positive
(negative) skewness in the initial density field results in larger
(smaller) halo concentrations. We crudely include this effect by
modifying the linear overdensity in equation (\ref{eq:colz}) as follows
\citep{matarrese00}: 
\begin{equation}
 \delta_c(z) \rightarrow \delta_c(z)\sqrt{1-S_3\delta_c(z)/3}.
\end{equation}
We estimate the skewness $S_3$ as $S_3\sim 6f_{\rm
 NL}\sigma_M^{-1}\sigma_\phi$ with $\sigma_\phi=4\times 10^{-5}$
\citep{dalal08}. In this model, primordial non-Gaussianity of
$f_{\rm NL}=\pm 200$ translates into the modification of median 
concentration parameters for haloes with the mass $10^{15}M_\odot$ 
by $\sim \pm 5\%$. 
Given the current level of constraints on $f_{\rm NL}$ from WMAP
\citep[e.g.,][]{komatsu03,spergel07,yadav08,hikage08}, in 
this section we consider three non-Gaussian models, $f_{\rm NL}=-100$,
$100$, and $200$, as well as the Gaussian case $f_{\rm NL}=0$ studied in
the previous sections.  For each model, we compute 300 realisations of
all-sky cluster catalogue to estimate median and cosmic variance of
large lenses. First, we examine the effect of primordial
non-Gaussianities on the probability distribution of the largest Einstein
radii $\theta_{\rm max}$. The dependence of $\theta_{\rm max}$ on
$f_{\rm NL}$ is displayed in Figure~\ref{fig:ein_ng}. We find that
primordial non-Gaussianity of $|f_{\rm NL}|\sim 100$ hardly affect
$\theta_{\rm max}$. Although $\theta_{\rm max}$ is increased by
$\sim 10\%$ from $f_{\rm NL}=-100$ to $200$, it is clearly smaller than the
cosmic variance. This suggests that the observation of $\theta_{\rm
max}$ hardly constrains $f_{\rm NL}$, at least not so tightly as the
current WMAP data do. On the other hand, the plot shown in
Figure~\ref{fig:sep_ng} suggests that we can in principle detect
primordial non-Gaussianities of $|f_{\rm NL}|\sim 100$ from the all-sky
abundance of clusters with relatively large Einstein radii,
$N(>\theta_{\rm E})$. For instance, $N(>20'')$ for $z_s=1$ is  $34\pm
6$ and $60^{+10}_{-8}$ (68\% confidence) for $f_{\rm NL}=0$ and $200$,
respectively. The abundances of large lenses for higher source redshifts
are more sensitive to $f_{\rm NL}$, because lens clusters are located at
higher redshifts where the cluster abundance is more sensitive to
primordial non-Gaussianities. One of the reasons why the abundance of
large Einstein radii is better in probing $f_{\rm NL}$ than the
observation of $\theta_{\rm max}$ is its smaller cosmic variance.

The results presented above suggest that constraints on $f_{\rm NL}$ are
not improved very much by including large lenses, compared with
current CMB constraints. However, some inflation models predict strongly
scale-dependent primordial non-Gaussianities \citep[e.g.,][]{loverde08},
and thus independent constraints from clusters of galaxies, which probe
smaller scales (a few Mpc) than CMB anisotropies ($\ga 100$~Mpc), can be
very important to test such  scale-dependence. The best constraints on
primordial non-Gaussianities at the cluster scale are expected to be
obtained by the number count of clusters at high-redshifts, detected in
radio, X-ray, or optical, but an accurate calibration of cluster masses
is always challenging \citep[e.g.,][]{hu07,takada07}. The statistics of
large lenses may therefore provide an important complementary test  of
cluster-scale primordial non-Gaussianities. 

\section{Discussions}
\label{sec:dis}

\subsection{Observational strategy}
\label{sec:survey}

In this paper, we derived all-sky distributions of large Einstein
radii based on the F$\Lambda$CDM model. An advantage of the Einstein
radius statistics is that it is determined quite well from 
observations, provided that strongly lensed arcs are observed.
How many arcs do we expect? \citet{oguri03} computed the
number of lensed arcs in a typical massive cluster, with a mass of
$2-3\times 10^{15}M_\odot$, to be $\sim 1$ for the arc magnitude limit
of $\sim 26$~mag. Since the clusters studied in this paper have $2-3$
times larger Einstein radius than typical clusters of similar masses,
the expected number of lensed arcs for these clusters should also be
larger by a factor of $5-10$. Therefore, we conclude that reasonably
deep ($\sim 26$~mag) optical imaging of clusters having large Einstein
radii should always reveal several strongly lensed arcs, which will be
sufficient to determine their critical curves accurately. This is indeed 
the case in the largest known lens clusters such as A1689
\citep{broadhurst05b}, CL0024+1654 \citep{kneib03,jee07},
RXJ1347$-$1145 \citep{bradac05,bradac08,halkola08}, SDSS~J1209+2640
\citep{ofek08}, and RCS2~2327$-$02 (M. Gladders et al., in
preparation) in which many strong lensing events are already
identified.   

The discussion above suggests that wide-field deep optical surveys, such
as done by the Large Synoptic Survey Telescope 
\citep[LSST;][]{ivezic08}\footnote{http://www.lsst.org}, 
provide a promising way to locate clusters with very large Einstein
radii. In such optical surveys, we will be able to search for strongly
lensed arcs directly without a priori information on locations of
massive clusters \citep[e.g.,][]{horesh05,seidel07}. Such
blind/automated arc survey has been attempted by \citet{cabanac07}
using  Canada-France-Hawaii Telescope Legacy Survey (CFHTLS) data. The
detection and characterisation of massive clusters using weak lensing
\citep[e.g.,][]{miyazaki07} will complement the identifications of
lensed arcs and will allow a check of the model of clusters that we
adopted in Section \ref{sec:mc}. Another approach is to make use of
(shallower) optical,  X-ray, or Sunyaev-Zel'dovich (SZ) surveys to
identify candidate massive clusters, and conduct follow-up optical
imaging of each massive cluster to characterise its lensing
properties. Examples of such optical/X-ray/SZ cluster surveys include
the maxBCG cluster survey \citep{koester07}, the ROSAT-ESO Flux
Limited X-ray (REFLEX) Galaxy 
cluster survey \citep{bohringer04}, the Massive Cluster Survey
\citep[MACS;][]{ebeling07}, the ROSAT PSPC Galaxy Cluster Survey
\citep{burenin07}, the Red-Sequence Cluster Survey
\citep[RCS;][]{gladders05}, and  planned SZ cluster surveys such as the
South Pole Telescope (SPT)\footnote{http://pole.uchicago.edu}, the 
Atacama Cosmology Telescope
(ACT)\footnote{http://www.physics.princeton.edu/act/}, the Atacama
Pathfinder EXperiment (APEX) SZ
survey\footnote{http://bolo.berkeley.edu/apexsz/}, and a survey using 
the Planck satellite.  However, clusters with largest Einstein radii
are not necessarily the most massive.  Cluster surveys need to be deep
enough to locate masses as small as $\sim 5\times 10^{14}M_\odot$ to
assure completeness (see also Figure \ref{fig:massz}).   

\begin{figure}
\begin{center}
 \includegraphics[width=0.85\hsize]{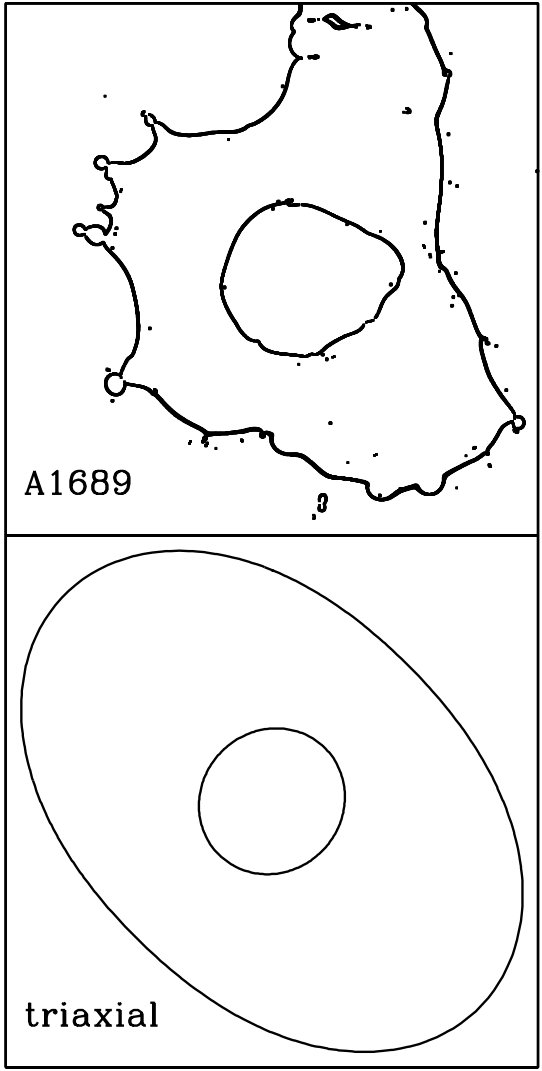}
\end{center}
\caption{Critical curves of A1689 ($z_s=3$) constrained from multiple
 strong lens systems \citep{broadhurst05b} are compared with the
 most plausible critical curves of the largest lens in our model,
 which is obtained from the median parameter values listed
 Table~\ref{table:ein}.  
\label{fig:crit_ze}}
\end{figure}

The critical curves of the largest lenses predicted in our model may
offer guidance for identifying such systems in observations. 
In Figure~\ref{fig:crit_ze}, we plot our prediction for the plausible
critical curves of the largest lens, as well as the critical curves of
A1689 obtained in \citet{broadhurst05b}. Because of the high
concentration and rounder shape of the projected mass distribution,
the predicted inner critical curves are rather small compared with the
outer critical curve. Therefore for these systems strong lens events are 
dominated by standard ``double'' and ``quadruple'' image configurations. 
This is in marked contrast to typical clusters in which less
concentrated 2D mass distributions increase the importance of their
inner critical curves and produce naked cusp image
configurations \citep[e.g,][]{blandford87,oguri04,oguri08}. We find
that the critical curves of the largest known lens cluster A1689 are
similar despite the perturbation by several substructures.  

One of our main findings is that large lens clusters represent a
highly biased population. Whilst the biases in projected 2D mass
distributions can directly be
tested from lensing observations of clusters, the alignment between
major axes and line-of-sight directions will require additional
observations. For instance, the line-of-sight elongation of a cluster
can be inferred by combining multi-wavelength data such as X-ray, SZ,
and kinematics of member galaxies 
\citep[e.g.,][]{fox02,lee04,gavazzi05,sereno07,ameglio07}.

\subsection{Effect of baryons}

The triaxial halo model of JS02, which we adopted, is based on $N$-body
simulations of dark matter. It is of interest to see how baryon
physics can affect our results. The most important baryonic effect on
cluster strong lensing comes from the central galaxy
\citep[e.g.,][]{meneghetti03b,sand05,puchwein05,hennawi07b,rozo08a, 
wambsganss08,hilbert08}.
Although the central galaxy can boost the lensing probability as high as 
$\sim 100\%$, the effect is clearly scale dependent such
that clusters with smaller Einstein radii are more substantially
affected by central galaxies. Both analytic and numerical studies
agree in that for lensing with large Einstein radii ($\theta_{\rm
  E}\ga 20''$) the enhancement of lensing probabilities due to central
galaxies become negligibly small \citep[see,
e.g.,][]{oguri06,wambsganss08,hilbert08,minor08}, suggesting that
the clusters discussed in this paper, which have unusually large
Einstein radii, are not affected by central galaxies. This is also
expected from the fact that their critical curves extend far beyond
central galaxies. Therefore we conclude that the effect of central
galaxies is negligibly small for our results. 

Baryons also influence the shape of clusters. For instance, dissipative
gas cooling results in more spherical dark haloes \citep{kazantzidis04}.
In addition, the inclusion of hot gas components slightly enhances the
concentration of dark haloes \citep{rasia04,lin06}, although current
studies are insufficient to quantify its statistical impact.

\subsection{Other possible systematics}

Cluster substructures affect the shapes and locations of arcs and
hence may influence the Einstein
radius. However, \citet{meneghetti07} found that the radial shift due to
substructures is $\la 5''$  \citep[see also][]{peirani08}, much less
than $\theta_{\rm E}$ considered in this paper, and the effect cancels
out to first order. 

Some clusters have quite complicated morphology which cannot be
described by a simple ellipsoid. An extreme example is the merger as
seen in the bullet cluster \citep{clowe06}, which
turned out to have a large impact on arc statistics \citep{torri04}. 
One of the reasons for the large effect of mergers on arc cross
sections is that complicated mass distributions 
tend to induce many cusps in caustics where prominent, long and thin
arcs are preferentially formed. In contrast, the size of the Einstein
radius is simply determined by the mass it encloses, and therefore the
Einstein radius should be less sensitive to the complexity of the mass 
distribution. In addition, clusters which undergo merger events can
be represented approximately by triaxial haloes with very small axis
ratios $a/c$. In this sense our calculation includes the effect of
mergers. The more robust treatment of mergers will
require calibration using high-resolution $N$-body simulations. 

In our calculations we have ignored any chance projection of multiple
clusters along the line-of-sight. We make a rough estimate for the
expected number of such events as follows. We consider following two
situations: (1) superposition of two massive haloes, and (2) a small
halo on top of a massive halo. For case (1), we require that both
haloes should have masses $M>7\times 10^{14}M_\odot$, because the
superposition of two clusters with these masses result in the total
Einstein radius of the system large enough to affect our results
($\theta_{\rm   E}\ga 30''$ for $z_s=3$). From the all-sky number of
such clusters, $N_{\rm clu}\approx 4,400$ for WMAP5, we compute the
chance probability as  $N_{\rm
  clu}^2\Omega^{-1}\pi\theta^2[1+\omega(\theta)]\approx 0.1\times
[1+\omega(\theta)] \la 1$ \citep[e.g.,][for the angular correlation 
  function $\omega(\theta)$]{brodwin07}, much 
smaller than the predicted abundance of large lenses (e.g., 
$N(>35'')\approx 35$ for $z_s=3$). For case (2), we consider
superpositions of massive haloes with $M>1\times 10^{15}M_\odot$ and
smaller haloes with $M>1\times 10^{14}M_\odot$ from the same reason,
i.e., the superposition of these haloes with typical concentrations
yields the total Einstein radius with its size comparable to that
discussed in the paper. The all-sky numbers, $1,100$ and $86,000$
respectively, suggest the chance probability of $\approx 0.5 \times
[1+\omega(\theta)] \la 1$, again smaller compared with the predicted
abundance of superlenses. Thus we expect that the effect of the
chance projection of multiple clusters is not significant. 

However, it has been argued that A1689 may have a possible secondary
peak along line-of-sight \citep{andersson04,lokas06,king07},
which implies that multiple clusters might in practice have some
impact on the statistics of superlenses. Ray-tracing in large-box
$N$-body simulations should provide an important cross-check of 
how important projections of haloes along line-of-sight are.

\section{Conclusion}
\label{sec:conc}

We have calculated the expected distributions of large Einstein radii in
all-sky (40,000~${\rm deg^2}$) using a triaxial halo model. 
Our approach to generate all-sky mock catalogue of massive clusters
and the properties of individual clusters with the Monte-Carlo method
allows us to evaluate the cosmic variance for such statistics, and at
the same time, to study biases in the population of clusters having
such large Einstein radii.   

The largest Einstein radius in all-sky for source redshift $z_s=1$
was predicted to be $42{}^{+9}_{-7}$ arcseconds for WMAP5,
$35{}^{+8}_{-6}$ arcseconds for WMAP3, and $54{}^{+12}_{-7}$
arcseconds for WMAP1, where errors are
$1\sigma$ cosmic variance. The sensitivity to $\sigma_8$ suggests that
this statistic is a good measure of it. The Einstein radii are
approximately twice as large for 
larger source redshift, $z_s=7$. In some realisations the largest lens
cluster is the most massive cluster in the universe; in others 
smaller than $10^{15}M_\odot$. We have found that the population of
these ``superlens'' clusters are significantly biased: their major
axes are almost always aligned with the line-of-sight, and their
projected 2D mass distributions appear rounder (by $\Delta e\sim0.2$)
and more concentrated ($\sim 40-60\%$ larger values of concentration 
parameters). These biases are stronger than those found in more common
lens clusters with smaller Einstein radii \citep{hennawi07b}. In
particular we have pointed out that the high concentration observed in
A1689 is consistent with our theoretical expectation at the $1.2\sigma$
level. Thus the combined analysis of several clusters will be
essential to address the claimed high concentration problem.
Finally, we have studied the effect of primordial non-Gaussianities,
and concluded that the abundance of relatively large lens clusters can
in principle constrain primordial non-Gaussianities at a level
comparable to the current CMB experiments ($|f_{\rm NL}|\sim 100$), if
other cosmological parameters are fixed. 

It will be very important to compare our analytic predictions with
ray-tracing in $N$-body simulations. In particular, the large
cosmological Millennium Simulation \citep{springel05} has sufficient
resolution to resolve the centres of massive dark haloes for strong
lensing studies \citep{hilbert07,hilbert08}. Although its small box
size ($500h^{-1}{\rm Mpc}$) does not allow predictions for all-sky 
distributions of Einstein radii, we can use these simulations to
validate and calibrate our semi-analytical model predictions. The
comparison of our results with those from $N$-body simulations will be
presented in a forthcoming paper. 

The statistics of large Einstein radii provide an important opportunity
to test the standard F$\Lambda$CDM paradigm, as it probes both the
high-mass end of the cluster mass function and central mass
distributions of massive clusters. The measurement of Einstein radii
is fairly robust, and future all-sky samples will soon be available 
to perform this study.

\section*{Acknowledgments}
We thank Tom Broadhurst for providing us the critical curve data of
A1689 and for helpful discussions. We also thank Steve Allen, 
Maru{\v s}a Brada{\v c}, Jan Hartlap, Phil Marshall, Masahiro Takada,
Kaoru Takahashi, and Keiichi Umetsu for discussions.  We are grateful
to an anonymous referee for many helpful suggestions. This work was
supported by Department of Energy contract DE-AC02-76SF00515 and by 
NSF grant AST 05-07732. 
 


\label{lastpage}

\end{document}